# Comprehensive systematic review into combinations of artificial intelligence, human factors, and automation

Reza Khani-Shekarab[1], Alireza khani-shekarab[2]

[1]Department of Civil engineering, Tehran University, Tehran, Iran

[2] Department of Industrial engineering, Tehran University, Tehran, Iran

E-mails: Khani_re@ut.ac.ir

**Abstract:**

Artificial intelligence (AI)-based models used to improve different fields including healthcare, and finance. One of the field that receive advantages of AI is automation. However, it is important to consider human factors in application of AI in automation. This paper reports on a systematic review of the published studies used to investigate the application of AI in PM. This comprehensive systematic review used ScienceDirect to identify relevant articles. Of the 422 articles found, 40 met the inclusion and exclusion criteria and were used in the review. Selected articles were classified based on categories of human factors and areas of application. The results indicated that application of AI in automation with respect to human factors could be divided into three areas of physical ergonomics, cognitive ergonomic and organizational ergonomics. The main areas of application in physical and cognitive ergonomics are including transportation, User experience, and human-machine interactions.

# 1 Introduction

Automation is being developed for different industries for different reasons such as improving safety, increasing efficiency, and reducing the workforce. For example in the automobile industry, automation is being developed to increase productivity, decrease production time, and reduce cost. Also, Autonomous features in cars are being developed to automatically control speed, park, and keep tracking within a lane [1]. Although the core idea behind automation is performing different actions independently from humans, almost all autonomous systems and subsystems need interaction with humans who act as controllers, supervisors, and the final responsible for system performance. Automation in most cases needs wide human monitoring, even though in many situations automation interfaces do not provide required data to the human operator [1]. Besides, developing effective automation systems is dependent on the following Human Factors and Ergonomics (HF/E) approach.

Human factors, also called ergonomics, is defined as *the application of psychological and physiological principles to the engineering and design of products, processes, and systems* [2]. The International Ergonomics Association (IEA) defines HF/E as, *the scientific discipline concerned with the understanding of interactions among humans and other elements of a system, and the profession that applies theory, principles, data, and methods to design to optimize human well-being and overall system performance* [3]. The HF/E field has developed to tested different theories and tools in order to ensure the well-being of human workers. [4]. Therefore, HF/E is concerned with the identifying of interactions between humans and elements of systems, and applying theory and methods to optimize system performance and human well-being [2]. The main objectives of HF/E are increasing productivity, enhancing safety and comfort, and reducing human error. This field is an integration of several disciplines including engineering, biomechanics, psychology, sociology, physiology, visual design, user experience, anthropometry, industrial design, interaction design, and user interface design.

In specific, the HF/E field is concerning with following items: designing processes, equipment, tools and that fit the human body; designing processes, equipment, tools and that fit the human cognitive abilities; Preventing repetitive strain injuries and musculoskeletal disorders; developing easy-to-use interfaces to equipment and machines; fulfilling the objectives of occupational health and safety and productivity; creating "fit" between the environment, equipment, and user [2]. Three domains of specialization of HF/E are including physical, cognitive, and organizational. Physical ergonomics is related to the activities, physical-elements, and interactions; Cognitive ergonomics is related to usability, human-computer interaction, and user experience engineering, and human mental processes such as perception, reasoning, memory, motor response; Organizational ergonomics is related to optimizing the surrounding socio-technical systems including organizational structures, processes, and policies as it is shown in Figure 1 [4].

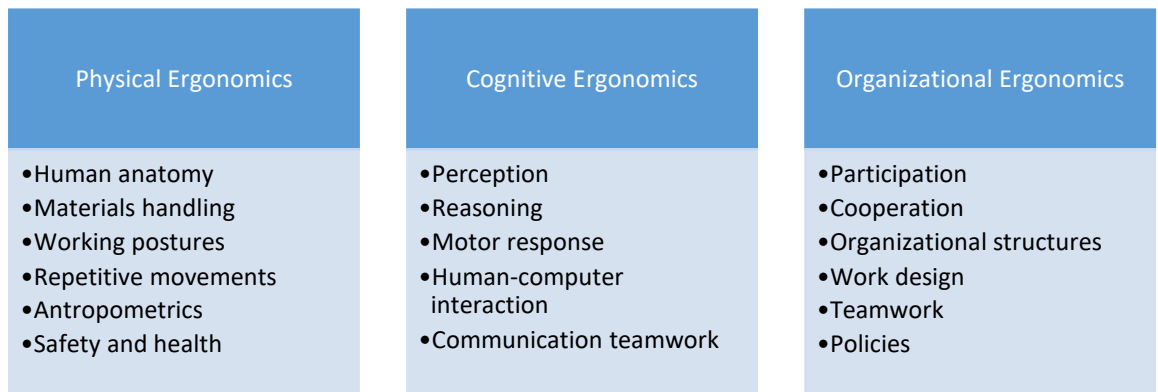

Figure 1. Different elements of Human Factors and Ergonomics

Lately, Artificial Intelligence (AI) has been proved effective in different fields of study. AI contains different computational technologies developed to learn from data and act reasonably [5]. Artificial intelligence (AI) can also help to overcome human factors challenges and improve successful automation [6]. Human factors and AI are two disciplines that followed almost parallel trajectories to improve the efficiency of work. They also both complement each other and overlap in various problem-rich domains. However, these two disciplines can be applied to improve each other; for example, Human factors can be used to improve interactions of AI-based models with users and in contrast, AI can be used to enhance different elements of human factors in different industries. AI in the context of human factors and automation can be used to (1) promoting autonomy, (2) predicting the human cognitive, (3) anticipate human physical states, (4) highly efficient in analyzing massive datasets of human measurements, and (5) enabling new human factors methods [7]. However, the main challenges are surrounding the application of AI in the context of human factors and automation such as narrow scientific understanding of the human brain, and providing appropriate training datasets [7].

The purpose of this review is to inform human factors researchers and practitioners with the latest improvement in the human factors issues related to automation in the context of artificial intelligence as is indicated in Figure 2.

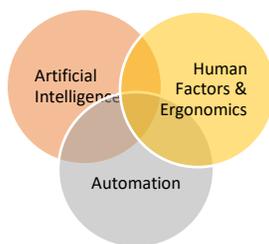

Figure 2. The objective of this review

## 2 Methodology

The Preferred Reporting Items for Systematic Reviews and Meta-Analyses (PRISMA) guidelines were used to select literature [8]. For this systematic review, two main features of research questions and search strategies were developed. The following research question considered:

- How artificial intelligence, human factors, and automation combined in published literature?

The search strategy contained different elements of (1) identifying main keywords and searching for relevant articles; (2) selecting the relevant articles; and (3) resolving the risk of bias among records[9]. A combination of three sets of keywords was defined to identify records.

- First set: Artificial intelligence, machine learning, pattern recognition, deep learning.
- Second set: human factors, ergonomics.
- Third set: automation, automatization, human-machine interaction

ScienceDirect was used to search for articles. 422 articles were identified and inclusion and exclusion criteria were used to screen the articles as it is represented in Figure 3. The inclusion criteria were articles written in English, and articles related to the objective of the article. The exclusion criteria were article in other languages, letters, newspaper articles, viewpoints, short papers, and posters.

The main biases in this methodology can happen through applying inclusion/ exclusion criteria, and/or categorizing included papers. To address these biases, the author selected included papers in several steps including reviewing abstract and conclusion, selecting relevant records, and reading the full text to select final records.

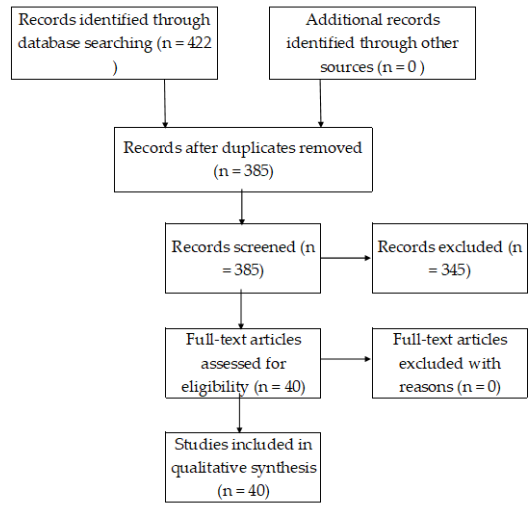

Figure 3. Chart of selection strategy following PRISMA guidelines.

## 3 Results

List of included papers and their categories and sub-categories are represented in Table 1.

Table 1. Included papers.

| References | Human Factors | Sub-category |
| --- | --- | --- |
| [10] | Cognitive Ergonomics | Data wrangling, exploratory analysis, and natural language translation |
| [11] | Cognitive Ergonomics | Product design assisting |
| [12] | Cognitive Ergonomics | Safe interactions |
| [6] | Cognitive Ergonomics | Relationship between AI and automation |
| [13] | Physical Ergonomics | Transportation |
| [14] | Physical Ergonomics | Transportation |
| [15] | Cognitive Ergonomics | Transportation |
| [16] | Physical Ergonomics | Transportation |
| [17] | Cognitive Ergonomics | Transportation |
| [18] | Physical Ergonomics | Work disability management |
| [19] | Physical Ergonomics | User experience evaluation |

| [20] | Physical Ergonomics | Evaluation of work tasks |
|---|---|---|
| [21] | Cognitive Ergonomics | Manufacturing tasks |
| [22] | Cognitive Ergonomics | planning human-robot shared tasks |
| [23] | Physical Ergonomics | Safe human-robot collaboration |
| [24] | Physical Ergonomics | Safety in shared work-space human-robot collaboration |
| [25] | Physical Ergonomics | Human-machine interaction for safety |
| [26] | Physical Ergonomics | Safe Industrial Robots |
| [27] | Cognitive Ergonomics | Virtual reality wearables |
| [7] | Physical and Cognitive Ergonomics | Machine Learning and Human Factors |
| [28] | Physical Ergonomics | Transportation |
| [29] | Organizational Ergonomics | Assessing continuous operator workload |
| [30] | Organizational Ergonomics | Cognitive workload estimation |
| [31] | Cognitive Ergonomics | Robotics collaborative task |
| [32] | Organizational Ergonomics | Supporting team Decision-making |
| [33] | Physical Ergonomics | Transportation |
| [34] | Organizational Ergonomics | Teamwork |
| [35] | Cognitive Ergonomics | Enhancing Brain-Computer interface |
| [36] | Organizational Ergonomics | Detecting non-verbal signals |
| [37] | Organizational Ergonomics | Organizational decision-making in multi-agent system |
| [38] | Cognitive Ergonomics | Virtual agents in explainable AI interaction design |
| [39] | Physical Ergonomics | Automation of software tests |
| [40] | Cognitive Ergonomics | Building trust |
| [41] | Cognitive Ergonomics | Building trust |
| [42] | Organizational Ergonomics | Heterogeneous industrial networks |
| [43] | Cognitive Ergonomics | AI-empowered Software Assistants |
| [44] | Physical Ergonomics | Transportation |
| [45] | Physical Ergonomics | Transportation |

| [46] | Physical Ergonomics | Transportation |
| [47] | Physical Ergonomics | Transportation |

# 4 Discussion

In this section, physical, cognitive, and organizational ergonomics among included papers are discussed.

## 4.1 Physical ergonomics

Recently, AI-based approaches have used to analyze physical ergonomics of products and services. Because of a significant correlation between workers' body postures and ergonomic risks, different partial and full body postural assessment approaches have been investigated. Recently, researchers practice using AI for developing smart postural analysis tools. It is reported that these tools can accurately assess the full-body postures of workers. For instance, a combination of AI and visual management approaches is used to develop a flexible full-body fuzzy-based postural evaluation tool [20]. The main part of this tool is creating a Fuzzy knowledge base approach including the core assessment rules of the suitable full-body assessment checklists. After investigating all postures, this tool can represent the ergonomic incidence for each posture.

Furthermore, AI-based approaches have used to improve physical ergonomics of products and services. Makki and associates (2019) discussed improving the performance and quality of Virtual Reality (VR) wearable devices in the rapidly growing market [27]. As Human Factors play an important role in improving the level of user acceptance, the study used AI to define Human Factors of VR wearable devices [27]. The paper proposed a supervised learning method, an adaptive multi-label classification model, to assess different aspects of human factors (wearability, safety, satisfaction, usability, and aesthetics) [27]. The results indicated that the usability aspect of human factors is the most important aspect of VR wearables devices following by wearability [27].

In physical ergonomics, AI-based systems can be used to predict different aspects of human factors such as safety. These systems have main parts of (a) identifying relevant elements human factors, (b) extracting and selecting features, and (c) developing a machine learning model. In this topics, Cheng and associates [18] developed a work injury database and used a fusion of Variational Autoencoder, Neural Turning Machines, and Long and Short Term Memory to predict accurate cost of work injuries, and provide advice on medical care.

Safety is one of the main aspects of physical ergonomics and human factors. Many researchers in automotive industry investigated application AI in human factors to improve safety. McDonald and associates (2014) developed an AI algorithm to detect drowsiness-related lane departures among drivers [28]. The study used steering wheel angle data and applied a random forest algorithm to detect driver drowsiness [28]. Another major safety issue of human factors in automotive industry is a driver distraction, which means drivers perform several additional tasks besides the essential driving tasks. Several studies integrated AI and human factors to develop the driving assistance system such as the customized user interface, non-distracting interface, and cognitive vehicle features [13,33].

Another aspect of physical ergonomics is user experience (UX) which is the users' behavior, perceptions, beliefs, emotions, preferences, physical and psychological responses that happen during, before, and after using a service or product. In this topic, AI models can be trained with user behavior monitoring data to better evaluate users' interactions in online systems [19].

## 4.2 Cognitive ergonomics

Due to constraints of human-machine interactions in aeronautics industry, there is a need to build new tools for analyzing and describing these interactions. Lately, researchers used AI-based systems to improve human factors in human-machine interactions. For instance, an object-oriented AI-based system can analyze human pilot activities and measure appraisal of pilot error, then it can offer a feedback to increase air safety [16]. The combination of AI and human factors has used in other areas of aeronautics industry. In safety of air traffic control domain, an AI-based system can be used to improve interactions between air traffic controllers and automated agents.

Advanced human-machine interactions play important function in self-driving cars. In this area, AI can bring a new form of human-machine interactions. For instance, several articles developed AI-based virtual assistance to help human in self-driving cars.

In military industry, a combination of AI and human factors can play important role in human-machine interactions. Because of advancement in developing military equipment, using AI-based technologies can improve human-machine interactions. Jie and associates (2017) developed a Multi-Mode military unmanned aerial vehicles (UAV) human-computer interaction (HCI) System by using artificial intelligence as its core part [14]. The study indicated that AI can be used in different functions such as eye-tracking, motion capture, voice recognition to improve response time, and mission performance of the UAV HCI system [14].

AI can be used to improve human-machine interactions in healthcare. Gnanayutham, (2003) used AI, fuzzy logic algorithm, to improve neurorehabiliatory communication interfaces for clinically brain damaged patients [35]. However, it is important to pay attention to impact of using AI on impact on automation bias and patient interaction [48].

One of the main aspects of human-machine interactions in industry is to calculate and monitor the trajectory of robots in real-time. In this matter, AI-based systems can be used to optimize productivity while considering safety requirements of humans. Calculating the trajectory of a robot can be performed in different steps including formulating safety constraints, determining of the kinematic configuration of a human around a robot by using sensor fusion algorithms, predicting safe space for the human worker, adjusting the pre-programmed trajectory, and executing safety constraints [22,23,26]. Monitoring the trajectory of a robot can be done by using computer vision, object recognition, and other AI-based algorithms to scan shared workplace between the human and robot [12,24,25].

One way to improve human-AI interaction is using AI to simulate real-time human-machine interactions. Matsas & Vosniakos, (2017) used AI techniques and developed an approach, Virtual Reality Training System, to (1) improve user experience inside the simulated world while interacting with the machine, and (2) determine acceptance of human-machine interactions and creating a reliable platform and guideline for human-machine interactions [21]. Another way to improve human-machine interactions is using a context-driven AI to improve team performance in human-machine interactions, and to increase situation awareness [31]. However, increasing integration of AI into human–machine interactions could lead to more complex cooperative human-machine systems. For example, it is reported that using AI can decrease controllability and predictability in cooperative systems [49]. Therefore, it is important to pay more attention to human factors in these complex systems [50].

Understanding emotions can play important role in efficient human-machine interactions. Different studies indicated that recognizing emotions could improve the human-machine system capability to predict the counterpart's intentions and social attitudes. In this area, extracting sequences of non-verbal signals and capturing the social function of emotions are investigated to understand emotions in multi-agent systems [36,37]. For instance, Dehais and associates (2012) mentioned the higher heart rate accompanied with attentional focus in humans can be an indicator of the occurrence of the conflict between the human operator and the robot [51].

### 4.3 Organizational Ergonomics

In organizational ergonomics, AI-based decision-aided systems can have positive or negative impacts on the organizational decision-making process. For instance, Jones (2015) investigated the impact of using an AI-based system on improvement of resource allocation decisions made by teams during a game [32]. The study indicated that teams with AI support had more errors than the teams without AI support. However, Canonico et al. [34] indicated that their AI-based model improved team-working. The proposed model was an integration of team cognition, collective intelligence, and AI.

Although AI has penetrated in many organizational decision-making processes, it is hard to imagine AI will soon replace humans in many decision-making processes [52]. In this area, Jarrahi [53] investigated the comparative benefits held by AI and humans concerning organizational decision making [53]. The author indicated that the main idea behind the partnerships between humans and AI is that AI can take care of mundane and ordinary works, while humans focus on creative tasks. Although humans have better intuitive and holistic approaches to handle uncertainty and equivocality, AI can offer superior analytical approach and more advanced computational information processing capacity to deal with complexity in organizational decision-making [53].

# 5. Conclusion

This systematic review is defined to better understand the combination of artificial intelligence, automation, and human factors. After defining the search question and search strategy, the keywords were used to identify the relevant records. The result indicated that the main aspects of human factors for application of artificial intelligence and automation are physical ergonomics, cognitive ergonomics, and organizational ergonomics. The main sub categories for selected papers are including human-machine interactions, trust in machine, user experience evaluation, and AI-empowered software assistants.